\documentclass[10pt]{article}
\usepackage{amsmath,amssymb,amsfonts}
\usepackage{units,hyperref,fancyhdr}
\usepackage{bbold,mathpazo}%opcoes de fonte: mathptmx, helvet, eulervm, avant, fourier, newcent, mathpazo
\usepackage{graphicx}% Include figure files
\usepackage{dcolumn}% Align table columns on decimal point
\usepackage{bm,bbm,mathtools,esvect}
\usepackage{slashed}
\usepackage{fullpage}
\title{\bf SELF-INTERACTING  QUANTUM  PARTICLES \\ AND THE DIRAC DELTA POTENTIAL \\  \vspace{5mm} }
\author{{\sf SERGIO GIARDINO\footnote{\tt sergio.giardino@ufrgs.br}}\\
\\
\small \it Departamento de Matem\'atica Pura e Aplicada \\
\small \it Universidade Federal do Rio Grande do Sul (UFRGS)\\
\small \it Caixa Postal 15080, 91501-970  Porto Alegre RS \\
\small \it Brazil}

\begin{document}
\date{}
\maketitle

\begin{abstract}
\noindent The Dirac delta function potential is considered within the real Hilbert space approach for complex wave functions, as well as quaternionic wave functions.  As has been previously determined, the real Hilbert space approach enables the possibility of self-interacting physical  systems. The self-interaction precludes confining states, and also imposes non-stationary quantum states, both of them representing novel situations that cannot be observed in terms of quantum wave functions. These results remark the differences between quaternionic quantum mechanics ($\mathbbm H$QM) and complex quantum mechanics ($\mathbbm C$QM), and also establish a method of solving the wave equation that may be applied to a variety of different cases.
\vspace{2mm}

\noindent {\bf keywords:} quantum mechanics; formalism; Scattering theory; other topics in mathematical methods in physics

\vspace{1mm}

\noindent {\bf pacs numbers:} 03.65.-w; 03.65.Ca; 03.65.Nk; 02.90.+p. 
\end{abstract}

\vspace{1cm}

\hrule
{\parskip - 0.3mm \footnotesize{\tableofcontents}}
\vspace{1cm}
\hrule

\pagebreak

%%%%%%%%%%%%%%%%%%%%%%%%%%%%%%%%%%
\section{INTRODUCTION\label{I}}%
%%%%%%%%%%%%%%%%%%%%%%%%%%%%%%%%%%

The mathematical generalization of quantum mechanics is the fundamental question behind the theoretical investigation presented in this article. Conceptually, a consistent generalized quantum mechanical theory in terms of quaternions  has been achieved in within the real Hilbert space \cite{Giardino:2018rhs}. However, this theory must be tested in several physical circumstances that will say whether such generalization is useful in order  to predict novel physical situations to be considered experimentally, as well as to reproduce well-known previous results.

In recent articles, this mathematical apparatus has been applied to the problem of the autonomous quantum particle \cite{Giardino:2024tvp}, where a self-interacting feature has been observed, and to examine the dynamics of expectation values as well \cite{Giardino:2025bym}. Conversely, the actual article extends these previous results and focus a further simple and important case, where the autonomous particle is exposed to the Dirac delta function potential. As in the previous article, the potential will be firstly considered in terms of a complex wave function, and then generalized to quaternionic wave functions. The  complex wave function is the correct precedent to serve as a reference to the quaternionic case, simply because the complex wave function is a particular case of the more general quaternionic wave function.

On the other hand, one must situate this research within the quaternionic applications to quantum mechanics. One will not give here the mathematical theory of quaternions ($\mathbbm H$), which can be found in several sources \cite{Ward:1997qcn,Dixon:1994oqc,Morais:2014rqc,Ebbinghaus:1990zah}, but only qualify them as non-commutative hyper-complex numbers that generalize the familiar complex numbers ($\mathbbm C$). 

From a physical standpoint, one can classify the quaternionic quantum applications according to the Hilbert space, which can be complex, quaternionic or real. In the usual complex Hilbert space, the applications do not generalize the theory, but comprise mathematical methods applied to a variety of quantum models, such the as wave equation  \cite{Arbab:2010kr,Graydon:2013sra,Sapa:2020dqm,Rawat:2022xsj}, monopoles \cite{Soloviev:2016qsx}, quantum states \cite{Steinberg:2020xvf}, the quantum concept of mass \cite{Arbab:2022cpe}, fermions \cite{Cahay:2019bqp,Cahay:2019pse}, and angular momentum \cite{deepka:2024nsf}, among many others. Within the quaternionic Hilbert space, one obtains the first attempt to generalize complex quantum mechanics ($\mathbbm C$QM), and this is made using a wave equation whose Hamiltonian operator is in fact anti-hermitian. Therefore, the quaternionic quantum mechanics ($\mathbbm H$QM) in the quaternionic Hilbert space is also called anti-hermitian. The pioneering book by Stephen Adler \cite{Adler:1995qqm} subsumes the huge volume of research within anti-hermitian $\mathbbm H$QM. Nevertheless, inevitable disadvantages affect this theory, first of all the absence of a well-defined classical limit ({\it c.f.} sec. 4.4 of \cite{Adler:1995qqm}), what implies the Ehrenfest theorem not to hold. A further serious drawback is the technically complicate form of the anti-hermitian $\mathbbm H$QM, where simple solutions are difficult to get mathematically, and to interpret physically as well. Despite of this, the several results concerning this subject include scattering \cite{Sobhani:2016qdp,Hassanabadi:2017wrt,Hassanabadi:2017jiz,Procopio:2017vwa,Sobhani:2017yee}, operators and potentials \cite{Ducati:2001qo,Nishi:2002qd,Madureira:2006qps}, wave packets \cite{Ducati:2007wp}, bound states \cite{DeLeo:2005bs,Giardino:2015iia}, perturbation theory \cite{DeLeo:2019bcw}, and quantum computing \cite{Dai:2023xxh}. 

Finally, the third possibility is the quaternionic quantum mechanics within the real Hilbert space \cite{Giardino:2018rhs}, a general theory that overcomes the problems of the anti-hermitian approach, encompassing a well-defined classical limit \cite{Giardino:2018lem}, as well as  simple analytical quantum solutions.  One accordingly recommends \cite{Giardino:2025bym,Rosa:2025git} as a detailed account of this theory, including the non-relativistic case, as well as the relativistic case, and a comprehensive record of previous work on this subject can be found within the same references. Therefore, the proposal of this article is to further examine the suitability of this theory in terms of the simple and well-known problem of the Dirac delta potential, and accordingly verify the theoretical power of the framework.
	
One also recognizes this article as an element within the debate concerning  the possibility of a real quantum mechanics ($\mathbbm R$QM), that contains favorable arguments \cite{Finkelstei:2022rqm,Chiribella:2022dgr,Zhu:2020iml,Fuchs:2022rih,Vedral:2023pij}, and also considerations against their viability \cite{Renou:2021dvp,Chen:2021ril,Wu:2022vvi,Li:2021uof,Batle:2024oqg,Weilenmann:2025yyg}. However, it is important to clarify that $\mathbbm R$QM  encompasses uniquely real quantum states in terms of a real Hilbert space because of the definition of the expectation value. Contrarily to this proposal, the  real Hilbert space $\mathbbm H$QM presented here is based on a definition of expectation value that allows complex and quaternionic wave functions within the  real Hilbert space, and thus conforming an alternative that unifies both of the ideas within a single theory.

After localizing the interest of this research within the general context of quantum theory, one must entertain the specific theme of the Dirac delta potential, which presents quantum significance on their own. One can mention recent applications within various themes, embracing non-linearity \cite{Ragan:2024mkx}, non-stationary scattering \cite{Chuprikov:2023dzk}, the Dirac delta potential within an infinite square well \cite{Girao:2023uwn},  fermionic interactions \cite{Santamaria-Sanz:2023fbp}, the modified Schr\"odinger spectrum \cite{Akbas:2023ucf}, Green's functions \cite{Castro-Alatorre:2022whh}, multiple delta potential \cite{Figueroa:2025lso}, point-like interactions \cite{kostenko20101,Granet:2022zpc}, the harmonic oscillator \cite{Ghose:2021vma}, super-symmetry \cite{MateosGuilarte:2014alt}, the $\delta'$ interactions \cite{Munoz-Castaneda:2014fxa,Fassari:2023twe}, two-particle system \cite{Deng:2024qjg}, and mathematical methods \cite{Gregory:2013ft,Kempf:2014nca,golovaty:2025neu}. Moreover, two review articles that present further research done in the field \cite{kostenko20101,Lange:2014ddf,Belloni:2014isq}. In summary, in this article one deploys the Dirac delta potential, whose relevance in quantum theory is already known, to reaffirm the viability of $\mathbbm H$QM in the real Hilbert space as a suitable generalized quantum theory. 

%%%%%%%%%%%%%%%%%%%%%%%%%%%%%%%%%%
\section{COMPLEX CASE}
%%%%%%%%%%%%%%%%%%%%%%%%%%%%%%%%%%

In this section, the well known quantum solutions for the Dirac delta potential will be obtained in an attempt to establish a mathematically general method to be applied in the quaternionic cases.
The study of the complex case is also required in order to demonstrate the absence of the self-interacting property that is exclusive of the quaternionic cases to be considered in the next section.
 The standard Schr\"odinger equation that a complex wave function $\psi$ must satisfy reads
\begin{equation}\label{dd01}
 i\hbar\frac{\partial \psi}{\partial t}=\left(-\frac{\hbar^2}{2m}\nabla^2+\mathcal V\right)\psi,
\end{equation}
provided $m$ as the mass of the particle submitted to the  scalar potential $\mathcal V$. Meanwhile, one must remember this investigation to be conducted within the mathematical groundwork of the real Hilbert space, and not within the usual complex Hilbert space. Henceforward, the expectation value of an operator $\widehat{\mathcal O}$ demands the definition
\begin{equation}\label{dd39}
 \big\langle\widehat{\mathcal O}\big\rangle =\frac{1}{2}\int d\bm x\left[\Psi^\dagger \widehat{\mathcal O}\Psi+\Big(\widehat{\mathcal O}\Psi\Big)^\dagger\Psi\right],
\end{equation}
where $\Psi^\dagger$ is the adjoint of the wave function $\Psi$. This definition differs from the current $\mathbbm C$QM mainly because it is evaluated over real numbers, but also because the quantum operator  $\widehat{\mathcal O}$ is arbitrary, and not necessarily of Hermitian character. Without the Hermitian requirement for physically relevant operators, $\mathbbm H$QM in the real Hilbert space  admits a wider range of operators as physically viable, and is evidently more general from this standpoint. 

After remarking these conceptual scenario, one can concentrate the attention to the specific scalar potential, accordingly chosen to be
\begin{equation}\label{dd04}
\mathcal V=q\delta(x)+V,
\end{equation}
whereby $\delta(x)$ is of course a Dirac delta function, and $q$ and $V$ are the complex constants
\begin{equation}\label{dd45}
 q=q_0+q_1i,\qquad \mbox{and}\qquad V=V_0+V_1i,
\end{equation}
defined in terms of the real quantities $q_0,\,q_1,\,V_0$ and $V_1$. In $x\neq 0$, where the Dirac delta does not act, it prevails the autonomous particle solution, composed of solutions such as
\begin{equation}\label{dd05}
 \psi(x,\,t)=A\exp\left[Kx-\frac{E}{\hbar}t\right],
\end{equation}
with the constant $A$ as the complex amplitude, and the complex parameters $E$  and $K$ as
\begin{equation}\label{dd46}
 E=E_0+E_1i,\qquad \mbox{and}\qquad K=K_0+K_1i.
\end{equation}
 The introduction of complex parameters in the scalar potential (\ref{dd45}) and in the wave function (\ref{dd05}) is a way to obtain more general wave functions in order to  describe non-stationary, and non-conservative phenomena. Within this picture, the complex quantities (\ref{dd46}) cannot be  identified as the energy and momentum parameters, and one must consider the expectation values in order to obtain these physical observables. One will briefly survey this subject in the sequel, and a complete description can be found in \cite{Giardino:2024tvp}.

\paragraph{AUTONOMOUS PARTICLE} The autonomous particle solution (\ref{dd05}) has been thoroughly considered in the real Hilbert space \cite{Giardino:2024tvp}, but one must remind it here as a means to entertain (\ref{dd04}). Substituting (\ref{dd05}) in (\ref{dd01}), wherever $x\neq 0$, it holds that
\begin{equation}\label{dd06}
 K_0^2-K_1^2=\frac{2m}{\hbar^2}\Big(V_0-E_1\Big),\qquad \mbox{and}\qquad 2 K_0 K_1=\frac{2m}{\hbar^2}\Big(V_1+E_0\Big).
\end{equation}
The first relation comes from the real part of the wave equation, while the second relation emerges from the imaginary component  and corresponds to a source of probability current, as described in terms of the continuity equation, as discussed in \cite{Giardino:2024tvp}. Intending to obtain the pair of solutions, and considering the parameters $E$ and $V$ as fixed, it emerges from (\ref{dd06}) the complete solution to be
\begin{equation}\label{dd08}
 \psi(x,\,t)=\Big(A_0\exp\big[Kx\big]+A_1\exp\big[-Kx\big]\Big)\exp\left[-\frac{E}{\hbar}t\right],
\end{equation}
where the amplitudes $A_0$ and $A_1$ are complex integration constants. Intending to obtain the physical characteristics of the autonomous particle, one has to determine physical expectation values.  In the real Hilbert space formalism \cite{Giardino:2018rhs},  and in agreement with the definition of the energy operator due to Schrödinger, the definitions of the energy operator, and the linear momentum operator announce
\begin{equation}\label{dd38}
 \widehat E=i\hbar\frac{\partial}{\partial t},\qquad  \mbox{and}\qquad
 \widehat p=-i\hbar\frac{\partial}{\partial x},
\end{equation}
and inevitably (\ref{dd39}) gives
\begin{eqnarray}
\nonumber&& \left\langle \widehat E\right\rangle=E_1\int\!\!\rho \,dx,\\
\nonumber && \Big\langle \widehat{ p}\Big\rangle=\hbar K_1 \int\!\!\rho\, dx,\\
\label{dd07}&& \left\langle \widehat{p^2}\right\rangle=\hbar^2\Big(K^2_1-K_0^2\Big) \int\!\!\rho\, dx,\\
\nonumber&&\left\langle \widehat V\right\rangle=V_0\int\!\!\rho\, dx,
\end{eqnarray}
wherein the probability density $\rho$ equals 
\begin{equation}
 \rho=|A|^2\exp\left[2K_0x-\frac{2E_0}{\hbar}t\right].
\end{equation}
Several physical interpretations arise from these outcomes. The physically observable quantities (\ref{dd07}) have a principal parameter, $E_1,\,K_1$ and $V_0$ for the total energy, linear momentum and potential energy, as well as $K_1^2-K_0^2$ for the kinetic energy. Additionally, all the quantities are modulated by $\rho$, which depends on $E_0$ and $K_0$. This common modulation factors out of the conservation of the energy relation
\begin{equation}
 \left\langle \widehat E\right\rangle=\frac{1}{2m}\left\langle \widehat p^2\right\rangle+\left\langle \widehat V\right\rangle,
\end{equation}
recovering the first relation of (\ref{dd06}), implying the conservation of the energy at every instant of the time variable, as well as establishing the relation between the principal parameters to hold  independently of time.
One also observes the admissibility of negative kinetic energies  whenever $K^2_1<K_0^2$ in (\ref{dd07}) provided  $E_1<V_0$ in (\ref{dd06}), an impressive quantum property that cannot be understood within the $\mathbbm C$QM in the complex Hilbert space.  Negative kinetic energies were firstly observed in \cite{Giardino:2025bym}, and are associated to non-stationary processes because $K_0$ generates a real exponential function responsible to change the amplitude of the wave function (\ref{dd08}). The inclusion of stationary and non-stationary processes within a single theoretical framework indicates the greater generality of the real Hilbert space formalism associated to the expectation value (\ref{dd39}), and is of course an exciting direction for future research.

Furthermore, the second relation of (\ref{dd06}) has been determined in \cite{Giardino:2024tvp} to be the continuity equation, associating non-conservative processes to the imaginary component $V_1$ of the potential and the real component $E_0$ of the complex energy. Particularly, $V_1$ contributes to specific physical situations, like inelastic scattering, and creation and annihilation of particles ({\it cf.} \cite{Schiff:1968qmq} Section 20).

A final analysis of the complex autonomous particle aims to reveal the conditions that identify stationary and non-stationary processes. One observes from (\ref{dd08}) that spatial and  temporal  coordinates  can be useful for describing non-stationary processes. In the case of the time coordinate, a pure imaginary $E$ generates a time stationary wave function, and thus $E_0=0$ is the required condition for temporal steady states. The spatial coordinate demands a more scrupulous evaluation.  Considering as initial conditions the complex parameters $E$ of the energy, and $V$ of the potential, the real components of the complex parameter $K$ accordingly are 
\begin{equation}\label{dd19}
 K_0^2=\frac{m}{\hbar^2}\left(V_0-E_1+\sqrt{(E_1-V_0)^2+(V_1+E_0)^2}\,\right)
\end{equation}
and 
\begin{equation}\label{dd20}
 K_1^2=\frac{m}{\hbar^2}\left(E_1-V_0+\sqrt{(E_1-V_0)^2+(V_1+E_0)^2}\,\right).
\end{equation}
Due to these values, steady state conditions can be determined, so that
\begin{equation}\label{dd09}
V_1+E_0=0\qquad \mbox{and}\qquad V_0<E_1, \qquad \mbox{implying that}\qquad K_0=0,\qquad \mbox{and}\qquad K_1\neq 0,
\end{equation}
thus meaning a pure oscillating stationary spatial behavior. Conversely, the pure non-oscillatory mode requires that
\begin{equation}\label{dd10}
V_1+E_0=0\qquad \mbox{and}\qquad E_1<V_0, \qquad \mbox{imply that}\qquad K_0\neq 0,\qquad \mbox{and}\qquad K_1= 0.
\end{equation}
Inversely,  $V_1+E_0\neq 0$ implies both of the real components of $K$ to be nonzero, and therefore along the spatial coordinate $x$ the wave function exhibits the oscillatory as well as the non-oscillatory behavior. In other words, either a forced or an attenuated oscillation. Nevertheless, this combined case is suitably classified as non-stationary.

\paragraph{DIRAC DELTA POTENTIAL I} To ascertain effect of the Dirac delta potential, one requires the integration of the whole Schr\"odinger equation (\ref{dd01}) for a particle submitted to the potential (\ref{dd04})  along an interval centered at $x=0$, so that
\begin{equation}\label{dd24}
 i\hbar\int_{-\epsilon}^\epsilon\frac{\partial \psi}{\partial t}dx=\int_{-\epsilon}^\epsilon\left(-\frac{\hbar^2}{2m}\frac{d^2\psi}{dx^2}+\mathcal V\psi\right)dx,
\end{equation}
in which $\epsilon>0$. The left hand side is identically zero, and the right hand side consequently produces
\begin{equation}\label{dd02}
\lim_{\epsilon\to 0}\frac{\hbar^2}{2m}\left[\frac{d\psi_+}{dx}(\epsilon,\,t)-\frac{d\psi_-}{dx}(-\epsilon,\,t)\right]=q\psi(0,\,t),
\end{equation}
where $\psi_+$ refers to the wave function within the region where $x>0$, and $\psi_-$ refers to the wave function within the region where $x<0$. The simplest physical case to be considered is
\begin{equation}\label{dd23}
\psi_-(x,\,t)=A\exp\left[Kx-\frac{E}{\hbar}t\right],\qquad \mbox{and}\qquad
\psi_+(x,\,t)=A\exp\left[-Kx-\frac{E}{\hbar}t\right].
\end{equation}
The wave function has to be continuous at $x=0$ as a condition to be a test function for the Dirac delta potential, so that
\begin{equation}\label{dd44}
\lim_{\epsilon\to 0-} \psi_-(\epsilon,\,t)=\lim_{\epsilon\to 0+} \psi_+(\epsilon,\,t)=\psi(0,\,t),
\end{equation}
and conditions (\ref{dd02}) and (\ref{dd44}) conclusively lead to
\begin{equation}\label{dd25}
 K=-\frac{mq}{\hbar^2}.
\end{equation}
 The complex quantity $q$ is a parameter that determines the linear momentum complex parameter $K$, and consequently establishes the physical character of the phenomenon as propagating in case of imaginary $K$, and non propagating in case of real $K$, as can be realized from the linear momentum expectation value in (\ref{dd07}). Moreover, from (\ref{dd25}) one immediately obtains the conservation of the energy relation
\begin{equation}
	E_1=\frac{m}{2\hbar^2}\Big(q_1^2-q_0^2\Big)+V_0,
\end{equation}
as well as the conservation of the probability relation
\begin{equation}
	E_0=\frac{m}{h^2}q_0q_1-V_1,
\end{equation}
remembering that these relations do not hold at the $x=0$ point because of the singular character of the Dirac delta function at this point. 
These simple results reveal physical features of the problem that are hidden in the usual complex solution, where 
\begin{equation} 
V_0=V_1=E_0=K_1=q_1=0.
\end{equation} 
First of all, one observes the usual negative energy term generated by $q_0$ is of kinetic origin, and not a contribution from the potential energy, as it is usually interpreted. The negative kinetic energy is a feature that can be understood only in terms of the real Hilbert space, and that cannot be understood in $\mathbbm C$QM. A further novel feature appears when the potential strength $q$ has non zero real and imaginary components, and either $E_0$ or $V_1$ have to be non zero. The wave function consequently describes a non-stationary physical state, a further  physical attribute of the problem that cannot be observed within the complex Hilbert space approach. One can clarify the interpretation using two simple particular cases.
 
\paragraph{PURE CONFINED STATES} The pure non-stationary case, which is the usual $\mathbbm C$QM solutions, requires condition (\ref{dd10}),  as well as (\ref{dd19}-\ref{dd20}), so that
\begin{equation}
K_0=\frac{\sqrt{2m(V_0-E_1)}}{\hbar},\qquad K_1=0,
\end{equation}
and the wave function assumes that
\begin{equation}\label{dd03}
\psi_-(x,\,t)=A\exp\left[K_0x-\frac{E}{\hbar}t\right],\qquad \mbox{and}\qquad
\psi_+(x,\,t)=A\exp\left[-K_0x-\frac{E}{\hbar}t\right],
\end{equation}
with complex amplitude $A$. The normalization of the wave function imposes $K_0>0$, and (\ref{dd25}) forces a real $q$, where $q_0<0$, and thus the delta potential represents a deeply infinite well. The energy parameter assumes
\begin{equation}\label{dd35}
 E_1=-\frac{m}{2\hbar^2}q_0^2+V_0,
\end{equation}
and the normalization requires
\begin{equation}
 A=\frac{\sqrt{mq_0}}{\hbar}.
\end{equation}
The correspondence between the real and complex approaches of the Hilbert space is exact in the limit $E_0=V_1=0$, but  the physical interpretation is rather different. Nonzero $E_0$ and $V_1$ generate novel solutions in terms of time dependent amplitude phenomena, and the real Hilbert space approach is free of the imaginary linear momentum expectation value that plagues the complex approach. Also the zero linear momentum expectation value is coherent to the non-propagating solutions of negative kinetic energy identified in \cite{Giardino:2024tvp}. 

\paragraph{SCATTERING STATES} The scattering states require $V_0<E_1$, and the  wave function reads
\begin{eqnarray}
\nonumber&&\psi_-(x,\,t)=\exp\left[Kx-\frac{E}{\hbar}t\right]+R\exp\left[\overline Kx-\frac{\mathcal E}{\hbar}t\right],\\
\label{dd43}&&\psi_+(x,\,t)=T\exp\left[-\overline Kx-\frac{\mathcal E}{\hbar}t\right],
\end{eqnarray}
where $R$ and $T$ are complex amplitudes, $\overline K$ corresponds to the complex conjugate of $K$,  and
\begin{equation}\label{dd48}
	\mathcal E=\mathcal E_0+E_1i.
\end{equation} 
The satisfaction of (\ref{dd06}) accordingly implies
\begin{equation}\label{dd47}
	\mathcal E_0+E_0=-2V_1,
\end{equation}
thus enabling normalizable wave functions. Condition (\ref{dd47}) relates the imaginary component of the scalar potential to time evanescent processes, a relation already observed in \cite{Giardino:2025bym}.
Condition (\ref{dd02}) and (\ref{dd44}) give the system of equations
\begin{equation}\label{dd37}
 1+R=T,\qquad  \overline KT+K+ \bar KR=\frac{2mq}{\hbar^2}T,
\end{equation}
and after defining the complex quantity
\begin{equation}\label{dd27}
 g=\frac{mq}{\hbar^2\overline K},
\end{equation}
the conditions (\ref{dd37}) consequently give
\begin{equation}\label{dd34}
 |R|^2+|T|^2=1+\frac{K_1}{\big|K\big|^2}\frac{2K_1+i\Big[\big(1+\bar g\big)K-(1+g)\overline K\,\Big]}{\big|1+g\big|^2}.
\end{equation}
The usual condition, namely real $q$, and pure imaginary $K$ recovers conservative process already known from $\mathbbm C$QM. However, the above result shows that non-conservative processes are admitted for complex $q$ and nonzero phase angles, something that cannot be understood in the usual complex Hilbert space approach. The research of physical phenomenon that admit such a description is a interesting direction for future investigation.

%%%%%%%%%%%%%%%%%%%%%%%%%%%%%%%%%%%%%%%
\section{QUATERNIONIC PARTICLES I}
%%%%%%%%%%%%%%%%%%%%%%%%%%%%%%%%%%%%%%%

In this section, one initially recalls the previous results of the autonomous particle \cite{Giardino:2024tvp} described by a quaternionic wave function 
\begin{equation}\label{dd12}
 \Psi=\psi_0+\psi_1 j,
\end{equation}
wherein $\psi_0$ and $\psi_1$ are complex functions, and $j$ an quaternionic imaginary unit. The basic elements of quaternions will not be considered here, and can be found elsewhere \cite{Ward:1997qcn,Dixon:1994oqc,Morais:2014rqc,Ebbinghaus:1990zah}. Nonetheless, the introduction of a quaternionic wave function (\ref{dd12}) is supposed to increase the capacity of describing physical phenomena due to the extra degree of freedom corresponding to $\psi_1$. However, quaternions impose algebraic constraints in virtue of the anti-commutative relation between the imaginary units
\begin{equation}
 ij=-ji,
\end{equation}
whose consequence
\begin{equation}
 \Psi i\neq i\Psi,
\end{equation}
imposes the existence of two workable wave equations, whose first alternative configures
\begin{equation}\label{dd13}
 i\hbar\frac{\partial\Psi}{\partial t}=\widehat{\mathcal H}\Psi.
\end{equation}
In the remaining possibility, the imaginary unit $i$ multiplies time derivative by the right hand side, and will be contemplated separately in the next section. The Hamiltonian operator $\widehat{\mathcal H}$, establishes
\begin{equation}
 \widehat{\mathcal H}=-\frac{\hbar^2}{2m}\nabla^2+\mathcal U,
\end{equation}
what generalizes (\ref{dd01}) by use of the quaternionic character of the scalar potential $\mathcal U$, in which
\begin{equation}\label{dd16}
\mathcal U=U_0+U_1j
\end{equation}
comprises the complex components
\begin{equation}
 U_0=V_0+V_1 i,\qquad \mbox{and}\qquad U_1=W_0+W_1 i,
\end{equation}
where   $V_0,\,V_1,\,W_0$ and $W_1$ are evidently real. The quaternionic wave function (\ref{dd12}) authorizes the decomposition of the quaternionic wave equation (\ref{dd13}) into a system of two coupled equations, virtually
\begin{eqnarray}
\label{dd14} && i\hbar\frac{\partial\psi_0}{\partial t}=-\frac{\hbar^2}{2m}\nabla^2\psi_0+U_0\psi_0-U_1\psi_1^\dagger\\
\label{dd15} && i\hbar\frac{\partial\psi_1}{\partial t}=-\frac{\hbar^2}{2m}\nabla^2\psi_1+U_0\psi_1+U_1\psi_0^\dagger,
\end{eqnarray}
wherein $\psi^\dagger$ corresponds to the adjoint of the complex function $\psi$.
Equation (\ref{dd14}) corresponds to the complex component of the wave equation, while (\ref{dd15}) necessarily
depicts the pure imaginary quaternionic component of the same equation. Supplementarily, one observes the pure imaginary quaternionic component $U_1$ of the scalar potential $\mathcal U$ to promote the coupling between the complex wave functions. This internal coupling can be understood as a self-interaction and is the relevant physical novelty presented by $\mathbbm H$QM in the real Hilbert space. 

As in the complex case, one intends to consider the Dirac delta potential in $\mathbbm H$QM, where the scalar potential (\ref{dd16}) becomes
\begin{equation}\label{dd11}
\mathcal U=Q\,\delta(x)+U,
\end{equation}
in which $Q$ and $U$ are quaternionic constants, so that
\begin{equation}
 Q=Q_0+Q_1j,\qquad \qquad U=U_0+U_1j
\end{equation}
and $Q_0,\,Q_1,\,U_0$ and $U_1$ are complex constants. In the region where $x\neq 0$, the solution corresponds to the wave function of a quaternionic autonomous particle, whose general solution has been systematically determined in \cite{Giardino:2024tvp} to be
\begin{equation}\label{dd17}
 \Psi=\big(A_0+A_1j\big)\exp\left[K x-\frac{E}{\hbar}t\right],
\end{equation}
in which $A_0$ and $A_1$ are complex amplitudes, as well as $K$ and $E$ are the  complex quantities
\begin{equation}
 E=E_0+E_1i,\qquad K=K_0+K_1i,
\end{equation}
whose real components accordingly are $E_0,\,E_1,\,K_0,$ and $K_1$. The expectation values, however, are slightly different from the complex scenario (\ref{dd07}), and then the inner product (\ref{dd39}), as well as the operators (\ref{dd38}) impose to the quaternionic picture that
\begin{eqnarray}
\nonumber&& \left\langle \widehat E\right\rangle=E_1\int\!\!\rho \,dx,\\
\nonumber && \Big\langle \widehat{\bm p}\Big\rangle=\hbar K_1 \int\!\!\rho\, dx,\\
\label{dd33} && \left\langle \widehat{p^2}\right\rangle=\hbar^2\Big(K^2_1-K_0^2\Big) \int\!\!\varrho\, dx,\\
\nonumber&&\left\langle \widehat V\right\rangle=V_0\int\!\!\varrho\, dx,
\end{eqnarray}
in which
\begin{equation}
 \rho=\Big(|A_0|^2-|A_1|^2\Big)|\psi|^2,\qquad \mbox{and} \qquad \varrho=\Big(|A_0|^2+|A_1|^2\Big)^2|\psi|^2,
\end{equation}
and then recovering the complex instance if $A_1=0$. As in the complex case in the real Hilbert space, this solution admits negative kinetic energy, non-stationary solutions. Nonetheless, the complex amplitudes are not independent, as will be seen in a moment. 

\paragraph{THE QUATERNIONIC AUTONOMOUS PARTICLE}
In this situation, $E$ and $\mathcal U$ are the free parameters, and $K$ has to be determined from a matrix equation provided by (\ref{dd14}-\ref{dd15}), and by the wave function (\ref{dd17}), so that
\begin{equation}\label{dd18}
\left[
 \begin{array}{cc}
   U_0+i E & -\, U_1\\
  \overline{ U}_1 & \overline{ U}_0- i\, E
  \end{array}
\right]\left[
\begin{array}{c}
 A_0 \\
 \overline A_1
\end{array}
\right]
=
\frac{\hbar^2K^2}{2m}
\left[
\begin{array}{c}
 A_0 \\
 \overline A_1
\end{array}
\right],
\end{equation}
remembering that the property
\begin{equation}
 \big(A_0+A_1j\big)\psi=A_0\psi+A_1\psi^\dagger j,
\end{equation}
has been used, and that $\,\hbar^2 K^2/2m\,$ wherein (\ref{dd18}) is the eigenvalue. The solution of this equation corresponds to the real quantities $K_0$ and $K_1$, obtained after defining
\begin{equation}
 \alpha=\Big(E_0+ V_1\Big)^2- E_1^2+ U_1\overline{U}_1\qquad \mbox{and}\qquad \beta=2 E_1\big( E_0+ V_1\big),
\end{equation}
so that
\begin{equation}\label{dd21}
 K_0^2=\frac{m}{\hbar^2}\left[V_0\pm\sqrt{\frac{\sqrt{\alpha^2+\beta^2}-\alpha}{2}}+\sqrt{\left(V_0\pm\sqrt{\frac{\sqrt{\alpha^2+\beta^2}-\alpha}{2}}\,\right)^2+\frac{\sqrt{\alpha^2+\beta^2}+\alpha}{2}}\;\right ],
\end{equation}
and 
\begin{equation}\label{dd22}
 K_1^2=\frac{m}{\hbar^2}\left[-V_0\mp\sqrt{\frac{\sqrt{\alpha^2+\beta^2}-\alpha}{2}}+\sqrt{\left(V_0\pm\sqrt{\frac{\sqrt{\alpha^2+\beta^2}-\alpha}{2}}\,\right)^2+\frac{\sqrt{\alpha^2+\beta^2}+\alpha}{2}}\;\right ].
\end{equation}
The complex components $A_0$ and $A_1$ of the eigenvectors of (\ref{dd18}) and are related as
\begin{equation}
 A_1=\frac{1}{\overline U_1}\left[-E_1\pm \sqrt{\frac{\sqrt{\alpha^2+\beta^2}-\alpha}{2}}-i\left(E_0+V_1\pm\sqrt{\frac{\sqrt{\alpha^2+\beta^2}+\alpha}{2}}\right)\right]\overline A_0.
\end{equation}
The stationary and non-stationary cases have to be determined from (\ref{dd21}-\ref{dd22}) in the same token as were determined for the complex case in  (\ref{dd19}-\ref{dd20}). The time stationary wave function requires a pure imaginary constant $E$, and $E_0=0$ is the required steady state condition. The  conditions for stationary states require
\begin{equation}\label{dd31}
 E_0=V_1=0,\qquad \mbox{and}\qquad E_1^2>V_0^2+U_1\overline U_1.
\end{equation}
implying the pure stationary case to be 
\begin{equation}\label{dd32}
K_0^2=0,\qquad \mbox{and}\qquad K_1^2=\frac{2m}{\hbar^2}\Big(\sqrt{E_1^2-U_1\overline U_1}-V_0\Big).
\end{equation}
Inversely, the pure non-stationary requires
\begin{equation}\label{dd42}
 K_0^2=\frac{2m}{\hbar^2}\Big(V_0-\sqrt{E_1^2-U_1\overline U_1}\Big), \qquad \mbox{and}\qquad K_1^2=0.
\end{equation}
For both of the cases, the amplitudes relate as
\begin{equation}
 A_1=\frac{E_1}{\overline U_1}\left(\sqrt{1-\frac{U_1\overline U_1}{E_1^2}}-1\right)\overline A_0.
\end{equation}
The quaternionic autonomous particles are thus characterized, and one can subsequently contemplate the simplest quantum solution for the quaternionic particle. After remembering these results from \cite{Giardino:2024tvp}, one considers the novel physical situation generated by the Dirac delta potential (\ref{dd11}).

\paragraph{THE QUATERNIONIC DIRAC DELTA POTENTIAL} 
With the aid of the wave function (\ref{dd17}), and their continuity at $x=0$, one has
\begin{equation}\label{dd40}
 \Psi_-=\mathcal A\exp\left[K x-\frac{E}{\hbar}t\right], \qquad \mbox{and}\qquad 
 \Psi_+=\mathcal A\exp\left[-K x-\frac{E}{\hbar}t\right],
\end{equation}
wherein $\mathcal A=A_0+A_1j$. As in the previous case, the wave function is continuous at $x=0$, thus compelling the energy $E$ and the quaternionic amplitude $\mathcal A$ to be are equal in each branch of the wave function (\ref{dd40}). The  integration of the wave equation around the point $x=0$, in analogy to (\ref{dd02}), generates
\begin{equation}\label{dd28}
\frac{\hbar^2}{2m}\left[\frac{d\Psi_+}{dx}(\epsilon)-\frac{d\Psi_-}{dx}(-\epsilon)\right]=Q\psi(0,\,t),
\end{equation}
what immediately gives
\begin{equation}\label{dd26}
 K=-\frac{m}{\hbar^2}\,\frac{\overline{\mathcal A}\,Q\,\mathcal A}{|\mathcal  A|^2},
\end{equation}
and thus recovers (\ref{dd25}) in the case of complex $\mathcal A$ and $Q$. Relation (\ref{dd26}) can be used to determine the components of $K$ using the system of complex equations (\ref{dd14}-\ref{dd15}), in which
\begin{equation}\label{dd29}
\left[
 \begin{array}{cc}
   Q_0 & -\,Q_1\\
  \overline{Q}_1 & \overline{Q}_0
  \end{array}
\right]\left[
\begin{array}{c}
 A_0 \\
 \overline A_1
\end{array}
\right]
=
-\frac{\hbar^2K}{m}
\left[
\begin{array}{c}
 A_0 \\
\, \overline A_1
\end{array}
\right].
\end{equation}
Considering (\ref{dd29}) as an eigenvalue equation, from the characteristic polynomial one obtains
\begin{equation}\label{dd30}
 K=-\frac{m}{\hbar^2}\Big(\mathfrak{Re}\big[Q\big]\pm\big|\mathfrak{Im}\big[Q\big]\big|i\Big).
\end{equation}
Comparing (\ref{dd26}) and (\ref{dd30}), 
\begin{equation}
 \frac{\overline{\mathcal A}\,Q\,\mathcal A}{|\mathcal  A|^2}=\mathfrak{Re}[Q]\pm\big|\mathfrak{Im}[Q]\big|i,
\end{equation}
which can be considered a very interesting property because it transforms the quaternion $Q$ into a complex using another quaternion $\mathcal A$. Additionally, (\ref{dd30}) presents a similarity to (\ref{dd25}), because the real and imaginary components of $Q$ and $K$ correspond in magnitude, and the physical interpretation is analogous to the complex particle, demonstrating the physical nature of the delta potential not to change between the complex and quaternionic cases. Moreover, from (\ref{dd29}) one finally obtains the relation between the complex wave amplitudes to be
\begin{equation}\label{dd41}
\overline A_1=\frac{1}{Q_1}\left(Q_0+\frac{\hbar^2}{m}K\right)A_0,
\end{equation}
something necessary in order to determine the expectation values (\ref{dd33}).  One can now consider the specific cases.

\paragraph{CONFINED STATES} This situation requires $K_1=0$, and (\ref{dd30}) imposes  $UQ$ to be  real. Therefore, the Dirac delta potential within equation (\ref{dd13}) is physically equivalent to the complex case, because in both of the cases the confined state is non-propagating. The sole difference to notice is the energy of the state, where
\begin{equation}
 E_1^2=\left(V_0-\frac{m}{2\hbar^2}\mathfrak{Re}[Q]\right)^2+U_1\overline U_1,
\end{equation}
obtained from (\ref{dd42}), replaces (\ref{dd35}). Therefore, the quaternionic case presents a richer structure, as expected.

\paragraph{SCATTERING STATES} Requiring (\ref{dd31}-\ref{dd32}), the  wave function reads
\begin{eqnarray}
	\nonumber&&\Psi_-(x,\,t)=\mathcal A\Bigg(\exp\left[Kx-\frac{E}{\hbar}t\right]+R\exp\left[\overline Kx-\frac{\mathcal E}{\hbar}t\right]\Bigg),\\
	&&\Psi_+(x,\,t)=\mathcal AT\exp\left[-\overline Kx-\frac{\mathcal E}{\hbar}t\right],
\end{eqnarray}
where $\mathcal A$ is quaternionic $R$ and $T$ are complex amplitudes, and $\mathcal E$ follows (\ref{dd48}). The continuity of the wave function at $x=0$ and (\ref{dd02}) give 
\begin{equation}\label{dd36}
 1+R=T,\qquad \overline KT+K+ \bar KR=\frac{2m}{\hbar^2}\,\frac{\overline{\mathcal A}\, Q\mathcal A}{|\mathcal A|^2} \,T. 
\end{equation}
In the same token as (\ref{dd26}), the left hand side of second equation above is complex and implies $\overline{\mathcal A}Q\mathcal A$ to be also complex. This rationale permits an exact analogy to the complex case (\ref{dd27}-\ref{dd34}), and the matrix formulation provides
\begin{equation}
\left[
 \begin{array}{cc}
   Q_0 & -\,Q_1\\
  \overline{Q}_1 & \overline{Q}_0
  \end{array}
\right]\left[
\begin{array}{c}
 A_0 \\
 \overline A_1
\end{array}
\right]
=
-\frac{\hbar^2}{2m}\frac{\overline KT+K+ \bar KR}{T}
\left[
\begin{array}{c}
 A_0 \\
\, \overline A_1
\end{array}
\right].
\end{equation}
This system can be solved as an eigenvalue equation (\ref{dd29}), and additionally using (\ref{dd35}), one transforms (\ref{dd36}) to
\begin{equation}
	1+R=T,\qquad  \overline KT+K+ \bar KR=\frac{2mq}{\hbar^2}T,
\end{equation}
where
\begin{equation}
 q=\mathfrak{Re}[Q]\pm\big|\mathfrak{Im}[ Q]\big|i.
\end{equation}
The equations thus reproduce the complex case (\ref{dd37}), and the physical interpretation is also identical, the sole difference concerns the values of the momentum parameter $K$, that has a dependence on the potential $U_1$ in the quaternionic case according to (\ref{dd32}). Therefore, the quaternionic potential representing the self-interaction alter the linear momentum, and the physical condition that determines the strength of the self-interaction is an interesting direction for future research.

%%%%%%%%%%%%%%%%%%%%%%%%%%%%%%%%%%%%%%%
\section{QUATERNIONIC PARTICLES II}
%%%%%%%%%%%%%%%%%%%%%%%%%%%%%%%%%%%%%%%
In this section, one considers the right quaternionic wave equation, the remaining alternative to (\ref{dd13}), that  of course reads
\begin{equation}
 \hbar\frac{\partial\Psi}{\partial t}i=\widehat{\mathcal H}\Psi,
\end{equation}
and consequently the wave function (\ref{dd12}) produces the complex system of equations
\begin{eqnarray}
&&\;\;\; i\hbar\frac{\partial\psi_0}{\partial t}=-\frac{\hbar^2}{2m}\nabla^2\psi_0+U_0\psi_0-U_1\psi_1^\dagger\\
&& - i\hbar\frac{\partial\psi_1}{\partial t}=-\frac{\hbar^2}{2m}\nabla^2\psi_1+U_0\psi_1+U_1\psi_0^\dagger.
\end{eqnarray}
Constant scalar potentials $U_0$ and $U_1$ and the wave function (\ref{dd17}), early considered in \cite{Giardino:2024tvp}, leads to
\begin{equation}
\left[
 \begin{array}{cc}
  U_0+i E & -\, U_1\\
  \overline{ U}_1 & \overline U_0+i\, E
  \end{array}
\right]\left[
\begin{array}{c}
 A_0 \\
 \overline A_1
\end{array}
\right]
=
\frac{\hbar^2K^2}{2m}
\left[
\begin{array}{c}
 A_0 \\
 \overline A_1
\end{array}
\right].
\end{equation}
The non-trivial solution, namely
\begin{eqnarray}
 && K_0^2=\frac{m}{\hbar^2}\left(V_0-E_1\pm\sqrt{\big(V_0-E_1\big)^2+E_0\pm\sqrt{V_1^2+U_1\overline U_1}}\right)\\
 && K_1^2=\frac{m}{\hbar^2}\left(E_1-V_0\pm\sqrt{\big(V_0-E_1\big)^2+E_0\pm\sqrt{V_1^2+U_1\overline U_1}}\right),
\end{eqnarray}
and self-interacting solutions require
\begin{equation}
 E_0\pm\sqrt{V_1^2+U_1\overline U_1}\neq 0,
\end{equation}
what means that pure stationary, as well as pure non-stationary results are not possible in the self-interacting case, where the scalar potential is quaternionic. One must notice that the expectation values require a different definition operators, and the imaginary unit multiplies the right hand side of the function, so that
\begin{equation}
 \widehat E\Psi=\hbar\frac{\partial}{\partial t}\Psi i,\qquad  \mbox{and}\qquad
 \widehat p\Psi=-\hbar\frac{\partial}{\partial x}\Psi i.
\end{equation}
These definitions impose a difference on the expectation values where the expressions (\ref{dd33}) hold  with the probability density
\begin{equation}
 \rho=\varrho=|\mathcal A|^2|\psi|^2.
\end{equation}

\paragraph{THE QUATERNIONIC DIRAC DELTA POTENTIAL} Using the wave function (\ref{dd40}), one obtains the same results as between (\ref{dd40}) and (\ref{dd41}). However, the right wave equation does not admit confined non propagating states because $K_1\neq 0$, as well as pure stationary states are also not allowed because $K_0\neq 0$. Therefore, only non-stationary processes are allowed, and no further discussion needs to be adduced, because it would be superfluous after considering the discussion of the previous quaternionic particle.

%%%%%%%%%%%%%%%%%%%%%%%%%%%%%%%%%%%%%%%%%%%%%%%%%%%%%%
\section{CONCLUSION}
%%%%%%%%%%%%%%%%%%%%%%%%%%%%%%%%%%%%%%%%%%%%%%%%%%%%%%

In this article, one studied the  Dirac delta  potential for complex and quaternionic wave functions, both of them within the real Hilbert space formalism. The results reproduce the well known $\mathbbm C$QM results in the complex Hilbert space, although explaining the confined states in terms of a negative kinetic energy, something that is not allowed in the standard theory. As a consequence, the quaternionic theory recovers every complex result for a Dirac delta potential as a limit case, demonstrating their higher generality. 

The presented outcomes likewise demonstrate the technical viability of the quaternionic theory, which was tested to the autonomous particle, and now to the Dirac delta potential. Of course, one has to entertain quaternionic versions of more sophisticated physical situations where well known quantum results have already been determined for complexes, and consequently ascertain further  differences between the quaternionic and complex formulations.

Nonetheless, the results of this article reinforces $\mathbbm H$MQ in the real Hilbert space as a theory of greater generality, which admits a self-interacting character that cannot be ascertained in the usual $\mathbbm C$QM. Subsequently,  the directions of future research are various, and include the generalization of every established result within the standard quantum theory.

\begin{small}
%\paragraph{ACKNOWLEDGMENT} The author gratefully thank the reviewer for various suggestions that significantly improved the manuscript.

\paragraph{DATA AVAILABILITY STATEMENT} The author declares that data sharing is not applicable to this article as no data sets were generated or analyzed during the current study.

\paragraph{DECLARATION OF INTEREST STATEMENT} The author declares that he has no known competing financial interests or personal relationships that could have appeared to influence the work reported in this paper.

\paragraph{FUNDING} This work is supported by the Funda\c c\~ao de Amparo \`a Pesquisa do Rio Grande do Sul, FAPERGS, grant 23/2551-0000935-8 within Edital 14/2022.
\end{small}

%%%%%%%%%%%%%%%%%%%%%%%%
%
%
%  BIBLIOGRAPHY
%
%

\begin{footnotesize}
%\bibliographystyle{unsrt} 
%\bibliography{bib_qdelta}

\end{footnotesize}
\end{document}